\documentstyle[aps,prl]{revtex}
\begin{document}
\twocolumn[ 
\preprint{\vbox{ \hbox{RUB-TPII-26/95} \hbox{hep-ph/9509283}}}
\title{Nucleon  Tensor Charges in the SU(2) Chiral Quark--Soliton Model }
\author{Hyun-Chul Kim$^1$, Maxim V. Polyakov$^2$
and Klaus Goeke$^1$}
\address{$^1$
Institute for  Theoretical  Physics  II, \\
P.O. Box 102148, Ruhr-University Bochum, \\
D--44780 Bochum, Germany  \\
$^2$ Petersburg Nuclear Physics
Institute, \\ Gatchina, St. Petersburg 188350, Russia}

\date{September 1995}
\maketitle
\widetext
\vskip -2.1in
\rightline{\vbox{ \hbox{RUB-TPII-26/95} \hbox{hep-ph/9509283}}}
\vskip1.6in
\begin{abstract}
We investigate the singlet $g_T^{(0)}$ and isovector $g_T^{(3)}$
tensor charges of the nucleon, which are deeply related to
the first moment of the leading twist transversity
quark distribution $h_1(x)$, in the SU(2)
chiral quark-soliton model.  With rotational $O(1/N_c)$
corrections taken into account, we obtain $g_T^{(0)}=0.69$ and
$g_T^{(3)}=1.45$ at a low normalization point of several hundreds MeV. 
 Within the same approximation and
parameters the model yields
$g_A^{(0)}=0.36$, $g_A^{(3)}=1.21$ for axial charges
and correct octet--decuplet mass
splitting.  We show how the chiral quark-soliton model interpolates
between the nonrelativistic quark model and the Skyrme model.
\end{abstract}
\draft
]    

\narrowtext

The complete information about the quark structure of the nucleon
in leading--order hard processes
is contained in three twist-two parton distributions.
Two of them $f_1(x)$ and $g_1(x)$ have been
studied extensively theoretically and measured in
deep--inelastic scattering experiments \cite{dis_review}.
The third transversity quark distribution $h_1(x)$
is unaccessible for measurements in inclusive deep--inelastic
experiments.  However the $h_1(x)$ plays an essential role in
polarized Drell--Yan processes \cite{RalstonSoper} and
other exclusive hard reactions\cite{h1_review,Collins,JaffeJi91}.
 The measurement of the $h_1(x)$ has been proposed recently by the RHIC spin
collaboration \cite{RHICproposal} and HERMES collaboration
at HERA \cite{HERMESproposal}.

The evolution equation for $h_1(x)$ has been
derived in refs.~\cite{Kodaira79,Artru}. Also it was shown by
Jaffe and Ji \cite{JaffeJi91} that the first moment of $h_1(x)$
is related to the nucleon tensor charge:
\begin{equation}
\int_0^1dx (h_1(x)-\bar{h}_1(x))=g_T^{f},
\label{first_moment}
\end{equation}
where $f$ is a flavor index ($f=u,d,s,\cdots$)
and the tensor charge $g_T^f$ is
defined as the forward nucleon matrix element~\cite{JaffeJi91,Ji_prep}:
\begin{equation}
\langle N| \bar{\psi}_f  \sigma_{\mu\nu} \psi_f |N \rangle = g_T^{f}
\bar{U}\sigma_{\mu\nu} U,
\label{tensor_charge_definition}
\end{equation}
where $U(p)$ is a standard Dirac spinor and
$\sigma_{\mu\nu}=\frac i2[\gamma_\mu, \gamma_\nu]$.
It is convenient to introduce singlet and isovector tensor charges:
\begin{equation}
g_T^{(3)}\;=\;g_T^u-g_T^d, \;\; g_T^{(0)}\;=\;g_T^u+g_T^d.
\end{equation}
 The tensor charges depend on the renormalization scale and
the corresponding anomalous dimension at one loop has been calculated
in refs.~\cite{RalstonSoper,Artru}: $\gamma=2\alpha_s/3\pi$.

Our aim is to calculate the tensor charges 
(\ref{tensor_charge_definition})
in the chiral quark--soliton model
($\chi$QSM, often called the semibosonized
Nambu---Jona-Lasinio model) at a low normalization point of several
hundreds MeV.

The $\chi$QSM has been successful in reproducing
the static properties of the baryons such as
the octet-decuplet mass splitting
\cite{DiPePo,Reinhardt,Meissner,BloDiaGoePetPobPark},
axial charges\cite{WakWat,BochumSPbOsakaTokio,BloPolGoe,BloPra}
and magnetic moments\cite{ChrPobNuclPhys,Kim1}
and their form factors~\cite{ChrPobNuclPhys,Kim2} (for details, see 
the recent review~\cite{ChrBloKimetal}).
The baryon in this model is regarded as a bound state of
$N_c$  quarks  bound by a non-trivial chiral field configuration.
Such a semiclassical picture of baryons can be justified in the
$N_c\rightarrow \infty$ limit in line with more general
arguments by Witten~\cite{Witten}.  
 A remarkable virtue of $\chi$QSM
is that the model interpolates between the nonrelativistic quark
model(NRQM) and the Skyrme model~\cite{Michal}.
In particular, due to such an interplay,
it enables us to examine
the dynamical difference between the axial and tensor charges
of the nucleon.

In the following, we employ the effective QCD partition function
from the instanton picture of QCD in the limit of low momenta.  It
is given by a functional integral over
pseudoscalar and quark fields~\cite{DyPe1}:
\begin{equation}
{\cal Z} \;=\; \int {\cal D}\Psi {\cal D}\Psi^\dagger
{\cal  D}\pi^A\:
\exp \left( i\int d^4x \Psi^\dagger iD\Psi \right) 
\label{Partfunc}
\end{equation}
where $iD$ and $U^{\gamma_5}$ denote the Dirac differential operator
and the pseudoscalar chiral field, respectively:
\begin{equation}
iD\;=\;\beta (- i \rlap{/}{\partial} + MU^{\gamma_5}+ \bar{m}{\bf 1}),
\;\;\;U^{\gamma_5}=e^{i\pi^A\tau^A\gamma_5}.
\label{Diracop}
\end{equation}
$\tau^A$ are Pauli matrices
and $M$ is the dynamical quark mass which arises as a result of the
spontaneous chiral symmetry breaking and is momentum-dependent. The
momentum dependence of $M$ introduces the natural ultra-violet cut-off
(inverse average instanton size $1/\rho \sim 600\ \mbox{MeV}$)
\cite{DyPe1} for the theory given by eq. (\ref{Partfunc})
and simultaneously brings a renormalization
scale to the model.   
The  $\bar{m}$ stands for the current quark mass
defined by $\bar{m} = (m_u+m_d)/2$ with isospin symmetry assumed.
The operator $iD$ is expressed in Euclidean space in terms of the
Euclidean time derivative $\partial_\tau$ and the Dirac one-particle
hamiltonian $H(U)$
\begin{equation}
iD\;=\; \partial_\tau \;+\; H(U)
\end{equation}
with
\begin{equation}
H(U)\;=\;\frac{\vec{\alpha}\cdot \nabla}{i}
\;+\; \beta MU \;+\; \beta \bar{m} {\bf 1}.
\end{equation}
One can relate the hadronic matrix element
eq. (\ref{tensor_charge_definition}) to a correlation
function:

\begin{equation}
\langle 0 | J_{B}(\vec{x},T)  \bar{\psi} \sigma_{\mu\nu} \tau^a
\psi J^{\dagger}_{B}(\vec{y},0)|0 \rangle
\label{corf}
\end{equation}
at large Euclidean time $T$ . The baryon current $J_B$ can be constructed
from quark fields:
\begin{equation}
J_B=\frac{1}{N_c!}\varepsilon^{i_1 \ldots i_{N_c}}
\Gamma^{\alpha_1 \ldots \alpha_{N_c}}_{II_3}
\psi_{\alpha_1 i_1}\ldots \psi_{\alpha_{N_c} i_{N_c}},
\end{equation}
where $\alpha_1 \ldots \alpha_{N_c}$ are spin--isospin indices,
$i_1 \ldots i_{N_c}$ are color indices, and the matrices
$\Gamma^{\alpha_1 \ldots \alpha_{N_c}}_{II_3}$ are chosen 
in such a way that the quantum numbers of the corresponding 
current are equal to $II_3$.  The correlation function 
(\ref{corf}) can be calculated in the effective
chiral quark model defined by eq.(\ref{Partfunc}) 
using $1/N_c$ expansion.  The related technique can be 
found in \cite{DiPePo,ChrBloKimetal,ja}.  Here we give a result 
for the tensor charges to the next to leading order of 
the $1/N_c$ expansion:
\begin{equation}
g_T^{(0)}\;=\; \frac{\alpha}{I}, g_T^{(3)}\;=\;  \beta + \frac{\delta}{I},
\label{delta}
\end{equation}
where $\alpha$, $\beta$, $\delta$ and $I \sim N_c$ are given by
\begin{mathletters}
\label{Eq:expression}
\begin{eqnarray}
\alpha&=& \frac{iN_c}{2}  \int d^3 x  \int \frac{d\omega}{2\pi}
\mbox {tr} \langle x | \frac{1}{\omega+i H}
\tau_i \frac{1}{\omega+i H}\gamma_5 \gamma^i|x \rangle , \\
 \beta&=& -\frac{N_c}{6} \int d^3 x \int \frac{d\omega}{2\pi}
 \mbox{tr} \langle x| \frac{1}{\omega + i H}
\tau_i\gamma_5 \gamma^i|x \rangle , \\
 \delta &=& \frac{iN_c}{6} \int d^3 x \int
\frac{d\omega}{2\pi}\int \frac{d\omega'}{2\pi}
{\cal P} \frac{1}{\omega-\omega'} \varepsilon^{ijk}\nonumber\\
&\times& \mbox{tr} \langle x | \frac{1}{\omega + i H}
\tau_i \frac{1}{\omega' + i H}\tau_j\gamma_5 \gamma_k|x \rangle , \\
 I&=& \frac{N_c}{2} \int d^3 x \int \frac{d\omega}{2\pi}
\mbox{tr} \langle x | \frac{1}{\omega + i H}
\tau_i \frac{1}{\omega + i H}\tau^i|x \rangle .
 \end{eqnarray}
\end{mathletters}
Having examined eqs.~(\ref{Eq:expression}a)--(\ref{Eq:expression}d)
in large $N_c$ limit, we find that $g_T^{(0)} \sim N_c^0$ and
$g_T^{(3)} \sim N_c$, which are
the same as in case of NRQM.  In NRQM the tensor charges are 
equal to the corresponding axial charges.
Though $N_c$ dependence of the tensor charges given above are
equal to that of the corresponding axial ones,
we shall show that the tensor and axial charges have a different
behavior in the limit of large soliton size (large constituent quark mass).

Before studying the tensor charges,
let us discuss how the surprisingly small value
of the singlet axial charge (so called `` {\em spin crisis}'')
is related to its asymptotic behavior in the limit of large
soliton size in the present model.  The
suppression in the ratio of the axial charges $g_A^{(0)}/g_A^{(3)}\sim
1/N_c$ in large $N_c$ limit does not
provide a solution to the ``{\em spin crisis}",
since NRQM shows the same $N_c$ behavior of the
singlet-isovector ratio but it simultaneously gives
$g_A^{(0)}=1$.  Hence, in order to understand the ``{\em spin crisis}",
it is necessary to seek an additional suppression
in the singlet-isovector ratio of axial charges.
 In the Skyrme model the ratio of the axial charges 
$g_A^{(0)}/g_A^{(3)}\sim 1/N_c^2$
is suppressed by the additional powers of the $1/N_c$
in comparison with NRQM \cite{Marek} which was suggested as 
a solution to the ``{\em spin crisis}".
However, this additional $1/N_c$ suppression  
is lifted in extensions of the Skyrme model by inclusion of 
vector mesons \cite{Park_Weigel}. In $\chi$QSM 
\cite{BloPolGoe,WakYosh} the ratio of axial charges is given by
$g_A^{(0)}/g_A^{(3)}\sim 1/N_c$ in contrast to the Skyrme
model. The difference is due to the non--locality of the effective
action for pions eq.~(\ref{Partfunc}) in $\chi$QSM. 
In other words, higher gradient terms  neglected in the Skyrme model
give a non-vanishing contribution to the singlet $g_A^{(0)}$
in large $N_c$ limit. 

$\chi$QSM interpolates between NRQM and the Skyrme one, {\em i.e.} 
in the limit of small soliton size  
it reproduces the results of NRQM, whereas in the opposite limit of
large soliton size it mimics the Skyrme model.
Besides the $1/N_c$ suppression, the ratio $g_A^{(0)}/g_A^{(3)}$ in our
model is quenched in the limit of large soliton size 
(large constituent quark mass)
by the inverse powers of the soliton size (quark mass).
Indeed the numerical calculations for the self--consistent soliton give
$g_A^{(0)} \approx 0.36$ \cite{BloPolGoe} which is a
relatively small number and is compared well with the experimental
value $0.31\pm 0.07$~\cite{EllKarl}.

Reviewing eqs.(\ref{Eq:expression}a)--(\ref{Eq:expression}d)
in the limit of large soliton size (large constituent quark mass), 
one can easily find that
$\alpha \sim (M R_0)^{2} $, $I \sim (MR_0)^3$ and
$\beta,\delta \sim M R_0$.  Therefore, the ratio of the tensor charges
$g_T^{(0)}/g_T^{(3)}\sim 1/(M R_0)^2$
is sizably reduced in the limit of large soliton size, while the
analogous analysis of the axial charges \cite{BloPolGoe,Michal} gives
even much stronger suppression in the ratio
$g_A^{(0)}/g_A^{(3)}\sim 1/(MR_0)^6$.  This observation
of the different behaviors between the axial and tensor charges
leads to a conclusion that the tensor charges might deviate from the
axial ones remarkably.

In the limit of $M R_0 \to 0$, $\chi$QSM
corresponds to NRQM and yields:
$g_T^{(0)}=g_A^{(0)}=1$,
$g_T^{(3)}=g_A^{(3)}=(N_c+2)/3 $ (derivation for the axial charges see
ref.\cite{Michal}).  Note that it is of great importance
to take into account the rotational $1/N_c$ corrections
($\delta$ contribution in eq.~(\ref{delta})) to
derive this result in $O(N_c^0)$ order.
The soliton in $\chi$QSM has a
radius $M R_0 \sim 1$, so that one could expect a deviation
from NRQM predictions as well as from the Skyrme model
results.  In  figure~1 we show the dependence of the tensor and axial
charges on the soliton size. The results were obtained by calculating the
functional traces in eqs.~(\ref{Eq:expression}a)--(\ref{Eq:expression}d)
according to the Kahana and Ripka method \cite{KaRi}
with a simple variational Ansatz for the profile function.
We take advantage of the inverse-tangent profile function
$P(r)=2 \mbox{ArcTan}(R_0^2/r^2)$ which has the correct
asymptotic behavior of the profile function at small and large distances.

 From figure~1 we observe that the axial and tensor
charges starting from the same values of $(N_c+2)/3\approx 1.67$ for the
isovector case and $1$ for the singlet one at small soliton size have
qualitatively different behavior for larger $MR_0$ --- the dependence of
the tensor charges on soliton size is weaker than 
the corresponding dependence of the axial charges. 
 This qualitative difference is in accordance with
the asymptotics of the charges in large soliton size:
\begin{eqnarray}
g_A^{(3)} &\sim& (MR_0)^2, \qquad  g_T^{(3)} \sim MR_0,\nonumber \\
g_A^{(0)} &\sim& \frac{1}{(MR_0)^4},  \qquad
g_T^{(0)} \sim \frac{1}{MR_0}.\nonumber
\end{eqnarray}
We see that indeed the asymptotic dependence of the tensor charges is
weaker than the corresponding dependence of the axial charges.
 From this one
can conclude that the tensor charges are closer to their values of
$g_T^{(0)}=1$ and $g_T^{(3)}=5/3\approx 1.67$
in NRQM than the corresponding axial charges.
The similar conclusions were obtained in the bag model \cite{JaffeJi91}.

In the above lines, we considered the dependence of $g^{(0)}_{A}$ and
$g^{(3)}_{A}$ (respectively $g^{(0)}_{T}$ and $g^{(3)}_{T}$) on $MR_0$.
This can be translated into a dependence on the Dirac radius $R_1$
and allows then a direct comparison with recent results of
Brodsky and Schlumpf~\cite{BrodskySchlumpf}.  For this we extracted
$R_1$ from our self-consistent calculation~\cite{Kim2} with several
constituent quark masses and plotted in the vicinity of the physical
point ($R_1=0.74\;\mbox{fm}$) $g^{(0)}_{A}$ and $g^{(3)}_{A}$ vs. 
$M_NR_1$ with $M_N$ being the proton mass. 
 We find that the slopes of these curves agree well with those 
of~\cite{BrodskySchlumpf}, though in ref.~\cite{BrodskySchlumpf}
the value of $g_A^{(0)}=0.6$ appears to be larger than
our $0.36$ and experimental value $0.31\pm 0.07$~\cite{EllKarl}.
It is interesting to note that those models having quite different
origins show comparable features. A detailed investigations will be 
presented elsewhere.   

In order to evaluate the tensor charges numerically, we employ
the self-consistent profile function obtained by diagonalizing
the Dirac hamiltonian in a box (we choose the radial box
size $D\simeq 10 \ \mbox{fm} $ to achieve good accuracy ) and
solving the self-consistent equations by iteration. The
technical details can be found in refs.~\cite{technical_review,tueb}.

We have calculated the tensor charges
for different values of the constituent
quark mass, which is the only free parameter of the model. The
corresponding results are reported in table~1.  
 As our preferred value of the constituent quark mass,
we choose $M=420$~MeV at which the model
reproduces with good accuracy many nucleon observables -- octet-decuplet
mass splitting \cite{BloDiaGoePetPobPark}, isospin splittings in baryon
 octet and decuplet \cite{BloPraGoePhysRev}, singlet axial charge
\cite{BloPolGoe,WakYosh}, magnetic moments, isovector axial charge
\cite{BochumSPbOsakaTokio} and
electromagnetic form factors \cite{ChrPobNuclPhys}.

Finally, we obtain:
\begin{equation}
g_T^{(3)}\approx 1.45, \qquad  g_T^{(0)}\approx 0.69,
\label{ten_num}
\end{equation}
or
\begin{equation}
g_T^{(u)}\approx 1.07, \qquad  g_T^{(d)}\approx -0.38.
\label{numval}
\end{equation}
We find that our results are close to those in the bag model
\cite{JaffeJi91} and consistent with QCD sum rule calculations
of refs.~\cite{Ji_prep,IofKho}.

It is worth noting that the dependence of the tensor charges on the
normalization point is rather weak:

\begin{equation}
g_T^{(f)}(\mu)=
\biggl(\frac{\alpha_s(\mu)}{\alpha_s(\mu_0)}\biggr)^{\frac{4}{29}}
g_T^{(f)}(\mu_0),
\label{evol}
 \end{equation}
as $\mu \to \infty $ the tensor charges slowly vanish. One can
use this equation to recalculate the tensor charges at higher
normalization points using the values of tensor
charges (\ref{ten_num}) at low normalization points.
 The value of normalizations point $\mu_0$ pertinent to our model
is not uniquely determined from first principles, one has to 
choose $\mu_0$ of the order of $\rho^{-1}\simeq 600~\ \mbox{MeV}$, 
but there may be a factor of order unity. To do this quantitatively 
we follow the approach of ref.~\cite{DiaPolWei} and define $\mu_0=a/R$ 
($R$ is average distance between instantons 
$\sim 1/200$~MeV$^{-1}$ ) with a dimensionless
parameter $a$ to be varied in the variational estimate of bulk properties
of the instanton medium. According to \cite{DiaPolWei} the parameter
$a$ can be varied without significant change of parameters 
of the effective low--energy theory eq.~(\ref{Partfunc}) from 
$a\simeq3$ to $a\simeq7$.  In this region of $a$ the one--loop
QCD coupling constant varies in region (see table~1. 
of ref.~\cite{DiaPolWei}): 
\begin{equation}
\frac{\alpha_s(\mu_0)}{2 \pi}=0.098 \pm 0.035.
\end{equation}
Using these numbers and evolution eq.~(\ref{evol}) one can estimate an
uncertainty of the tensor charge at high normalization points
due to the uncertainty in the determination of the
low normalization point $\mu_0$ pertinent to our model:
\begin{equation}
\frac{\Delta g_T(Q^2)}{g_T(Q^2)}\simeq \frac{4}{29}\cdot
\frac{\Delta \alpha_s(\mu_0)}{\alpha_s(\mu_0)}=0.05.
\end{equation}
 From this analysis 
we see that owing to the weak dependence of the tensor charges on the
normalization point our results eq.~(\ref{numval}) for the tensor charges
at low energy normalization points acquire an additional error of about 
$5\%$ being evolved to high normalization points.

We thank the referee for drawing our attention to a comparison of our
study with that of ref.~\cite{BrodskySchlumpf}.
This work has partly been supported by the BMBF, the DFG
and the COSY--Project (J\" ulich).  The work of M.P. is supported
in part by grant INTAS-93-0283.

\begin{table}
\caption{The tensor charges of the nucleon $g^{(0)}_{T}$ and
$g^{(3)}_{T}$ as varying the constituent quark mass $M$.}
\begin{tabular}{cccc}
$M$ &$370\; \mbox{MeV}$
& $420 \;\mbox{MeV}$
& $450 \;\mbox{MeV}$  \\  \hline
$g^{(0)}_{T}$ &0.756&0.688&0.686    \\
$g^{(3)}_{T}$ &1.446&1.449&1.466  \\
\end{tabular}
\end{table}

\begin{center}
{\Large Figure caption}
\end{center}

\noindent
{\bf Fig. 1}: The dependence of the axial and tensor charges on the
soliton size. 
The solid curve represents the $g^{(3)}_{T}$, while the
dashed curve draws the $g^{(3)}_{A}$.  The dot-dashed curve depicts the
$g^{(0)}_{T}$, whereas the dotted curve illustrates the $g^{(0)}_{A}$.
The small arrows stand for the values of $g^{(3)}_{T}=g^{(3)}_{A}=5/3$ and
$g^{(0)}_{T}=g^{(0)}_{A}=1$ in NRQM, respectively.  The large
arrows denote NRQM and Skyrme limit of the present model.
The constituent quark mass for this figure is
$M=370\;\mbox{MeV}$ to be consistent with ref.[24].
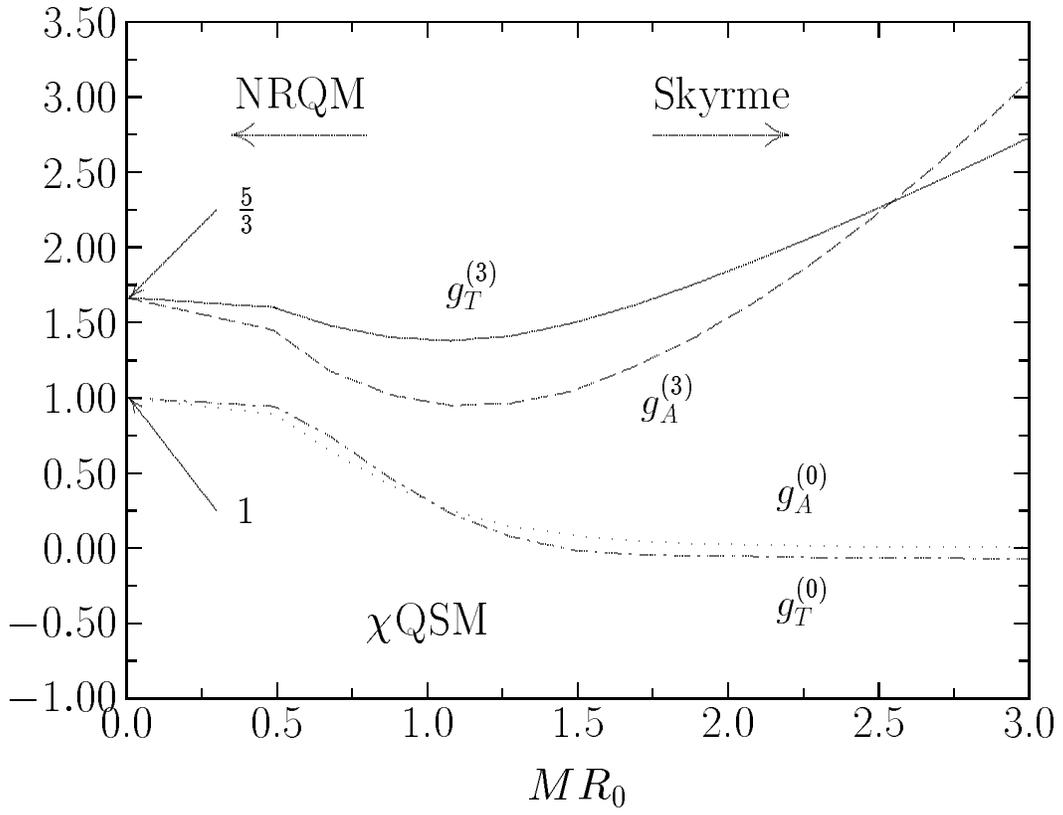
\end{document}